\documentstyle[12pt]{article}

\begin{document}

\newcommand{\bib}{\bibitem}
\newcommand{\er}{\end{eqnarray}}
\newcommand{\br}{\begin{eqnarray}}
\newcommand{\be}{\begin{equation}}
\newcommand{\ee}{\end{equation}}
\newcommand{\epe}{\end{equation}}
\newcommand{\bea}{\begin{eqnarray}}
\newcommand{\eea}{\end{eqnarray}}
\newcommand{\ba}{\begin{eqnarray}}
\newcommand{\ea}{\end{eqnarray}}
\newcommand{\epa}{\end{eqnarray}}
\newcommand{\ar}{\rightarrow}

\def\r{\rho}
\def\D{\Delta}
\def\R{I\!\!R}
\def\l{\lambda}
\def\D{\Delta}
\def\d{\delta}
\def\T{\tilde{T}}
\def\k{\kappa}
\def\t{\tau}
\def\f{\phi}
\def\p{\psi}
\def\z{\zeta}
\def\ep{\epsilon}
\def\hx{\widehat{\xi}}
\def\na{\nabla}
\begin{center}

{\bf Unified field theory from one-particle physics.}

\vspace{1.3cm}
M. Botta Cantcheff\footnote{e-mail: botta@cbpf.br}  

\vspace{3mm}
Centro Brasileiro de Pesquisas Fisicas (CBPF)

Departamento de Teoria de Campos e Particulas (DCP)

Rua Dr. Xavier Sigaud, 150 - Urca

22290-180 - Rio de Janeiro - RJ - Brazil.
\vspace{5mm}

\end{center}

\begin{abstract}

This work starts with the observation of a certain "rule" (up to now unexplored) 
 in the fundamental laws of Nature. We show some evidence of this, and formulate it
  as a fundamental principle which exhibits a number physical consequences.
   In particular, a new, very simple and extremely aesthetic unified model, which
    includes supersymmetry and supergravity, naturally arises from this principle,
	 together with some new "physics".
   
Furthermore, the new interpretation of Kaluza-Klein extra dimensions
  we advocate here provides a natural argument for 
 dimensional reduction, and the agreement with the observed phenomenology is recovered.
  In the high energy
  regime, a new physics is expected.

 Consequences in QFT are shortly commented. Finally, we observe a structure
  of "levels" and formulate a general conjecture about such a concept.
\end{abstract}
\section{Introduction}
We present here a remarkably simple unified theory of fields (including gravity)
 motivated (or inspired) by a hypothetized regularity of the natural laws.
 
In many aspects, the structure of (classical) field theory shares a great "similarity"
 with Classical Mechanics; for instance, the structure of the Nambu-Goto action,
\be
S_{NG}=\int (-g)^{1/2} \partial_a\phi^A \partial^a\phi_A,\label{sft}
\ee
for the "matter fields" in field theory (FT) is very similar to the action
 for a single particle in Classical Mechanics
\be
S_p = \int dt (dx^{\mu}/dt)^2. \label{sopt}
\ee
We realise this similarity by means of the correspondence $t \to M$ and $x \to \phi$.
 we claim that this is not a simple coincidence, but it rather reveals a
  fundamental fact of Nature with strong consequences.

We begin by stating the {\it single} fundamental hypothesis of the present work:
\\
\\

{\bf FT-OPT:} There exists a {\it universal} correspondence between the theoretical
 structure
 of FT and the One-Particle Theory (OPT). In particular, the world-line of a particle
  embedded in a spacetime $M$ corresponds to the embedding of $M$ in the
   {\it meta}-spacetime, ${\cal M}$, of the matter-fields of FT \footnote{FT is a {\it
    meta-theory} of one particle. This resembles a sort of "fractal"
	 behavior; we shall come back to this point at the end.}.
	\\
	\\
	
With the help of known facts of OPT, we find important ones for FT; for instance, SUSY and
 SUGRA.

In particular, a simpler and new unified model arises naturally from this principle. However, its
 general validity is not necessary in its formulation, and this model can be proposed
  independently from the fundamental hypothesis (FT-OPT).

\section{Direct consequences of FT-OPT in FT.}

We assume here some well-known points of the structure of OPT; they are listed below: 

OPT1) One-particle is a one-dimensional membrane, which moves on a 4-d Lorentzian manifold $M$.

OPT2) Its equation of motion is such that it describes a "geodesic" on $M$. This derives from
(\ref{sopt}), which corresponds to the "length" of the world line in $M$.

OPT3) The theory satisfies the full requirements that build up the kinematical structure of
 General Relativity (GR); for instance, covariance with respect to "general transformations 
of coordinates".

OPT4) The metric, $g$, of $M$ satisfies the Einstein Equation (E-E). 

These assumptions have their counterparts in FT, in agreement with the FT-OPT hypothesis,
namely:

FT1) The spacetime, $M$, is a d-dimensional membrane embedded in a {\it meta}-manifold,
 ${\cal M}$, the space of "matter-fields" \footnote{In agreement with known FT, this
  must be a complex manifold.}.

FT2) The equation of motion of $M$ corresponds to the minimal world
 volume ("d-dimensional geodesics"); they are the field equations.

FT3) The theory describing the background ${\cal M}$ is a "meta"-GR. 
In particular, we have two very useful facts of FT:
 
I. If we assume the existence of fermionic matter fields together with the bosonic one,
 general covariance
 (GC) requires SUSY. The meta-spacetime ${\cal M}$ is the {\it superspace}.

II. GC implies {\it local gauge symmetry}, but it has a richer structure: In
 Section 3, this is discussed in more details.

FT4) From GC with respect to transformations
 between the commuting and anti-commuting coordinates of ${\cal M}$,
 the metric, $G$, of ${\cal M}$ satisfies Super E-E (SUGRA).
 
Then, we also know: 
\be
D:=dim_{(bosonic)}[{\cal M}] < 12. \label{dim}
\ee


Notice that the {\it power} of FT-OPT: FT1...FT4, which are true matters in FT, have been
 remarkably obtained from it and from well-known facts of OPT. 

Another important remark is about the interpretation of the superspace:
 there are several (recent) embedded models \cite{ads} which
  work with this, but the interpretation of this space is different: this is the spacetime
   in itself, and the observed 4-d is typically obtained by Kaluza-Klein-type mechanisms
    \cite{witten}.
    In this approach, ${\cal M}$-space is interpreted
 as the space of matter fields -as we have mentioned above-, whereas the physical space-time
is some ($d<D$)-surface embedded in it, which {\it parametrizes} the evolution of the
 fields \cite{pst}.

 The dimensionality of this surface remains unexplained.
 

\section{Unification from FT-OPT.}

Notice that FT1.... FT4 already constitute the elements of a unified FT model.
\\
\\

"Our unified FT is defined by taking these ones to be {\it the fundamental assumptions}."
\\
\\

 We shall describe this in some more details.

 All the matter fields, $\phi^A$, (they might include fermions) play the
  same role as the coordinates  of a single particle in a Einstein's spacetime,
   and the background coordinates will be like the proper time; the resulting
    equation will correspond to the "minimal surface" in the (meta)space ${\cal M}$.
	 A minimal manifold is the natural generalization of the "geodesical hypothesis".
	  The embedding field $\phi^A (x^{\nu}): M \to {\cal M}$ describes the evolution
	   of those matter fields.

For simplicity, let us restrict ourselves to the bosonic sector of coordinates. Thus,
 the action must be:

\be
S := d m \int_M (-g)^{1/2} = m \int_M (-g)^{1/2} [q^{c}_a q^{d}_b G_{cd} g^{ab}], \label{action}
\ee

where $m$ is a fundamental constant and $q^{b}_a$ is the "projector" from $T_p{\cal M}$ into $T_p M$\footnote{Recall
 that in a Lorentzian general manifold $d = Tr(g)$.}, which may be written in terms of
  the embedding:
\be
q^{b}_a = [\partial_{\mu}\phi^A ] d_a x^{\mu} \frac{\partial^{b}}{\partial\phi_A}.\label{proy}
\ee

 $x^{\mu}$ denotes coordinates in the
 basis-manifold $M$ and $\phi^A$ are coordinates in ${\cal M}$. The latin indices $a, b$
  stand for the abstract ones \cite{wald} of ${\cal M}$, while the greek $\mu, \nu..$ and
   the capital $A, B...$ correspond to the coordinate frame of $M$ and ${\cal M}$,
    respectively. The covariant derivative is, according to FT3, compatible with $G_{ab}$.

	From (\ref{proy}), 
 it is easy to see that (\ref{action}) adopts the more familiar Nambu-Goto form \footnote{For $G$ flat,
  (\ref{action}) reduces to (\ref{sft}).}. 

We shall prove the agreement of this model with what is known for FT. This shall be done
 with the techniques
 of Dimensional Reduction (DR), but in a spirit remarkably different from the Kaluza-Klain
  (KK) picture.
 
Now, we are left with the task of showing this for interacting fields, and
 later for the matter-sector. 
 
Firstly, it is very important to have in mind the concept of Dimensional Reduction (DR).

The condition for Dimensional Reduction \footnote{DR-condition.} is that there exists a
 manifold $M$, embedded in ${\cal M}$, such that {\it every field on ${\cal M}$ is nearly
  function of the coordinates of $M$}, that is to say; if $f$ is some field on ${\cal M}$,
   then
\be
f \sim f(x^{\mu}).\label{dr}
\ee
such a manifold ${\cal M}$, is called {\it reducible manifold}.

In this framework, DR is naturally {\it ruled} by the energy of the system;
 the main reason is that the {\it extra} coordinates have a clear interpretation:
	
	 It can be seen, from action (\ref{action}), that if the energy is limited,
	  the matter-field amplitude, $|\Delta \phi|\le |(\Delta \phi)_{max}|$,
	   is bounded too (recall that they
  are associated to the $D-d$ coordinates). Then, natural units of $\phi$
 must combine to produce a constant $l$ such that $l\delta\phi$ has
  unit of length; if $l$ is small enough \footnote{This could be ruled by the
   fundamental constant $m$.}, a very large $|(\Delta \phi)_{max}|$ is need
   to observe some variation of $f$ \footnote{Any field of the theory.} with
    respect to $\phi$.
	
	"For small matter fields (low energy), we have DR-condition (\ref{dr})".

{\bf Interacting fields.}

Notice that, in principle, we have no gauge fields. From FT3, all the field theory is
 encoded in the metric $G_{ab}$, which satisfies the E-E (the bosonic sector). The field
  equation is:
\be
R_{ab}[G]-(1/2)G_{ab}R={\cal T}_{ab},
\ee
where ${\cal T}_{ab}$ is the energy momentum tensor, derived from the Lagrangian term
 (\ref{action}) in the usual way\footnote{In the contribution to the energy-momentum tensor,
  there should be a distribution on $M$ -proportional to $m$-.}.

Thus, it is evident that we have a {\it new physics} when DR-conditions do not hold. i.e,
 corrections to the current YM equations appear and then: {\it a new phenomenology might be
  expected when the amplitudes of matter fields are not negligible}.

Now, we shall show how to make contact with the observed interacting-field theory which
 is successfully enough described by an Einstein-YM theory\footnote{The standard model.}.

In a neighborhood of a point $ p\ep M$, the structure of
 ${\cal M}$ is $\sim M \times {\cal F}$.

For the interactions, the DR-scheme (KK-model) works whenever the following DR-condition holds:

 The metric $G$ at the point $q \ep {\cal M}$ depends only of the projection map of $q$
 into M: 
\be
[{\cal L}_{v} G_{ab}]|_{p \ep M} \sim 0,\label{kill}
\ee
 for {\it every} $v \ep T_p{\cal F}$ -this means that $v$ is a Killing vector in a neighborhood of $p$-.

 In other words, the dependence of $G$ on the fields $\phi^A$ can be neglected. And again,
  this occurs when the energy of the matter fields is low.

Using (\ref{kill}), the components of $G$ can be separated and identified
 with the M-metric $g$, and the 1-form gauge potentials; thus, we can find Einstein-YM
  theory,
  with the gauge group being that of the standard model, in the same
   way as doing dimensional reduction. As it has been shown by Witten \cite{witten},
    this requires $dim[{\cal M}]=11$ in remarkable agreement
   with the constraint (\ref{dim}).

{\bf On the gauge theories.}

Typically, the structure of the fiber ${\cal F}$ is considered linear, a (natural)
 representation space for the gauge group; but, this theory implies a {\it stronger
  locality} for the gauge fields, namely, the parameter $\alpha$ of a local gauge
   transformation
   not only depends on the spacetime point $x^{\mu}$, but also on the matter
    fields $\phi$; actually, it is a function of the point in ${\cal M}$. A meta-local gauge
	 transformation is actually a pointwise coordinate
	 transformation of ${\cal M}$. We naturally have corrections to YM equations for the
	  gauge
	  fields.

It is well-known that GR can be formulated as a gauge theory for the group of local
 coordinates transformations. In the present context, the current gauge theories (standard
  model) are built by restricting to "particular" diffeomorphisms of ${\cal M}$; the
   "gauge-coordinate transformations (GCT)":
\be
x^{\mu}\to x'^{\mu}=x'^{\mu}(x^{\nu}),
\ee
\be
\phi\to \phi'=u(x^{\nu})\phi(x^{\nu}),
\ee
where $u.u^{\dag}= 1$.
Notice that in this type of transformations, the local transformation-matrix is fairly
 independent from the matter fields, if $\alpha$ represents a matrix element, 
\be
\alpha(x,\phi) \sim \alpha(x), \label{var}
\ee
in agreement with DR-condition!. So, {\it GCT are the diffeomorphisms consistent with DR},
 and DR-condition could be implemented at this level, i.e FT could be built from the
  invariance with respect to GCT, as is well known. The current Einstein-YM is recovered
   when DR-conditions hold.

This theory is a {\it meta}-GR theory: this is a {\it meta-local} gauge theory, in the
 sense that (\ref{var}) does not hold and the gauge transformation is {\it field-dependent}.

Incidentally, field-dependent gauge transformations appear very often in SUSY and SUGRA in
 connection with Wess-Zumino-type gauges \cite{wz}.

{\bf Matter sector.}

FT1 prescribes the {\it structure} of FT but does not pick out the physical fields of
 the theory i.e, this does not specify which are the the coordinates of ${\cal M}$
 that represent the matter fields. Unfortunately, such a freedom provides us with various perspectives
 to find the correct
coupling between the matter and the gauge fields. Here we consider only one of these.
 
For simplicity, take $D=d+1$\footnote{Where the extra coordinate is assumed to be complex.}.
Starting with the action (\ref{action}), consider the particular embedding ($x^{\mu}\to \phi^A$):
 
\be
\phi^A= \left(\phi^{d+1}:=\phi(x^{\nu});\phi^{\mu}(x^{\nu})\right),
\ee
such that,
\be
\phi^{\mu}(x^{\nu}):= - i e\int \phi^{*}(x^{\nu}) dx^{\mu}, \label{K}
\ee
where $e$ is the "coupling" constant and $\phi$ represents the physical matter field. Then, we have:
\be
\partial_{\mu} \phi^A = (\partial_{\mu} \phi; - i e \phi^{*} \delta_{\mu}^{\nu}).\label{tranf}
\ee
The metric components, $G_{AB}$, are supposed to
 be nearly independent from $\phi^{d+1}$.

Thus, action (\ref{action}), written in these coordinates, reads as below:
\be
S_{NG}= m\int_M (-g)^{1/2} g^{\mu\nu}\left(G_{d+1, d+1} \partial_{\mu} \phi \partial_{\nu} \phi^{*} - i e \phi^{*} G_{\nu, d+1}\partial_{\mu} \phi + c.c.
- e^2 \phi \phi^{*} G_{\mu \nu}\right),\label{sft2}
\ee
which, with the K-K ans\"atze, {\it agrees} with the action for a
 charged scalar matter field \cite{salam}. The non-Abelian case is rather different and 
 shall not be analyzed here.

{\bf Quantization.}

Clearly, we can apply FT-OPT to the quantization, in such a way that the one-particle
 Quantum Mechanics corresponds to QFT. But, now, remarkably enough, QFT is the
  quantization of a {\it unified} FT -which includes gravity-, where the meta-background
   ${\cal M}$ would remain fixed. Thus, a QG would be defined.
 

\section{A conjecture about the possible {\it fractality} in the natural laws.}
 If
 we define {\it meta-fields} (fields on the meta-spacetime), we would have a meta-FT (MFT)
  too. 
 Thus, in principle it appears natural to go one step further in this context by applying
  again FT-OPT, establishing a correspondence between FT and MFT. So, we would have an
   interesting structure of the natural laws; a sort of {\it fractality}. We could
    postulate this as a fundamental fact, but the fundamental reason for these {\it jumps}
	 and their structure are mysterious and they
    should be investigated more accurately.

\section{Concluding remarks.}

Firstly, we need to stress again that the hypothesis FT-OPT is not required for the validity
 of this unified FT; the latter could be established by itself. {\it Nevertheless}, if we start
  from this FT, the hypothesis FT-OPT is remarkably satisfied.
  
In this framework, multiple {\it conceptual unifications} have appeared naturally:

- Matter fields and spacetime coordinates appear at the same conceptual level. This
 open up the possibility to analyze the space-time geometry in terms of 
{\it particles}\footnote{loosely speaking.}; clearly, the inverse also holds through.

- Susy, coordinate and gauge transformations are particular classes of the most
 general diffeomorphisms of ${\cal M}$.

- The space-time and the target metrics (appearing for instance, in strings,
 sigma models, and others) are the same entity.

  The energy-density of the matter fields is associated to the "amplitude of variation"
 of the non spacetime ($D-d$)-coordinates. Thus, energetic reasons for DR can naturally
 be argued.

Other issues have been solved as the old problem of the interpretation of the extra dimension
 in the Kaluza-Klein models; here, they are recognized as the matter fields. 

The surprising agreement with the dimensionality (D=11) required for a realistic FT must be
  remarked \cite{witten}.

In a forthcoming paper, we shall exploit physical consequences arising from this
 new formulation of FT.

A novel possibility has been put in: to interpret the fifth coordinate (or extra
 dimensions, in general) of the recent brane-world-type models \cite{ads}
  as a {\it field} on the 4-d brane, which is interpreted as the physical space-time.

\section{Acknowledgments.}

Thanks are due to J. A. Helay\"el-Neto for discussions, encouragement and a critical reading
 of the manuscript in its earlier form. CNPq is also
acknowledged for the invaluable financial help.


\begin{thebibliography}{99}
\bibitem{ads} Randall, L. and Sundrum, R., {\it Nucl.Phys. B557 (1999) 79};
 Maldacena, J. {\it Adv. Theor. Phys. 2 (1998) 231};
 Witten, E., {\it Adv.Theor.Math.Phys.2 (1998) 253}. 
\bibitem{witten} Witten, E. {\it Nucl. Phys. B186 (1981), 412.} And references therein.
\bibitem{pst} M. Botta Cantcheff, {\it"The 'phenomenological' space-time and quantization." , gr-qc/9804084}.
\bibitem{wald} Wald, Robert M. "General Relativity" (1984),
Chicago and London.
\bibitem{wz} De Witt, B. and Freedman, D. {\it Phys. Rev. D12 (1975), 2286.}
\bibitem{salam} Salam, A. and Strathdee, J. {\it Ann. Phys. (1982), 316.}
\end{thebibliography}
\end{document}